%% file: root.tex
\documentclass[3p,times]{elsarticle}

\usepackage{ecrc}
\volume{00}

\firstpage{1}

\journalname{\textcolor{black}{Applied Mathematics and Computation}}

\runauth{}

\jid{procs}

\CopyrightLine{2023}{Published by Elsevier Ltd.}

\usepackage{amssymb}
\usepackage[figuresright]{rotating}
\usepackage{lineno}
\usepackage{booktabs}      
\usepackage{graphicx}      
\usepackage{natbib}        
\usepackage{physics}       
\usepackage{amsmath}
\usepackage{color}
\usepackage{amsthm}
\usepackage{algorithm}
\usepackage[noend]{algpseudocode}
\usepackage{comment}
\usepackage{mathtools}
\usepackage[title]{appendix}
\usepackage{subcaption} 
\usepackage{caption}
\usepackage{xcolor}
\newtheorem{definition}{Definition}
\newcommand{\todo}[1]{\textcolor{red}{\textbf{\underline{TODO:}} #1}}
\newtheorem{theorem}{Theorem}[section]

\newtheorem{example}{Example}[section] 
\newtheorem{lemma}{Lemma}[section] 

\begin{document}

\begin{frontmatter}

\dochead{}

\title{A New Hybrid Automaton Framework with Partial Differential Equation Dynamics}

\author[1]{Tianshu Bao}
\author[2]{Hengrong Du} 
\author[3]{Weiming Xiang} 
\author[1]{and Taylor T. Johnson}

\address[1]{Institute for Software Integrated Systems, Vanderbilt University, Nashville TN 37240, USA}
\address[2]{Department of Mathematics, Vanderbilt University, Nashville TN 37240, USA}
\address[3]{School of Computer and Cyber Sciences, Augusta University, Augusta, GA 30912, USA}

\input{abstract}


\end{frontmatter}

\input{intro}
\input{PDHA}
\input{DSPDHA}
\input{example}
\input{conclusion}

\bibliographystyle{elsarticle-num}
\bibliography{bibliography.bib, Dubibliography.bib}

\end{document}

%% file: abstract.tex
\begin{abstract}
This paper presents the syntax and semantics of a novel type of hybrid automaton (HA) with partial differential equation (PDE) dynamic, partial differential hybrid automata (PDHA). 
In PDHA, we add a spatial domain $X$ and harness a mathematic conception, partition, to help us formally define the spatial relations. 
While classically the dynamics of HA are described by ordinary differential equations (ODEs) and differential inclusions, PDHA is capable of describing the behavior of cyber-physical systems (CPS) with continuous dynamics that cannot be modelled using the canonical hybrid systems' framework. 
For the purposes of analyzing PDHA,  we propose another model called the discrete space partial differential hybrid automata (DSPDHA) which handles discrete spatial domains using finite difference methods (FDM) and this simple and intuitive approach reduces the PDHA into HA with ODE systems.  
We conclude with two illustrative examples in order to exhibit the nature of PDHA and DSPDHA.
\end{abstract}

%% file: intro.tex
\section{Introduction}\label{sec:intro}
Cyber-Physical Systems have the potential to revolutionize the development of technology worldwide, and in recent years, numerous researchers and companies have invested significant time and money into ensuring that these systems will realize their economic and societal potential \cite{lee2008cyber}. However, since CPS fundamentally integrate computational and physical processes, there are unique challenges in designing and analyzing these systems.  In fact, the functional correctness of CPS relies deeply on the dynamics of their physical environment and the discrete control decisions of their computational units.  Within hybrid systems, the framework of HA has demonstrated considerable utility in capturing the complex interactions between the discrete and continuous parts of a CPS. Additionally, it allows for a formal analysis of the safety and reliability of CPS \cite{krishna2015hybrid}. However, this modelling framework has catered to systems with dynamics that can be described by ODEs, and hardly any attention has been paid to systems with dynamics described by PDEs \cite{tran2018adhs}. 

Meanwhile, safety of certain CPS relies on reachable set/state estimation results that are based on Lyapunov functions analogous to stability \cite{xiang2015equivalence, xiang2016necessary, xiang2018parameter, xiang2017stability, xiang2017robust, zhang2016mode, xu2016reachable} and reachability analysis of dynamical systems \cite{xiang2017output, xiang2017reachable},
certainly have potentials to be further extended to safety verification. One can study the stability of switched systems, a switched system is composed of a family of continuous or discrete-time subsystems along with a switching rule governing the switching between the subsystems, with the help of the given switching rules described by a prescribed state space partitioning \cite{johansson1997computation, margaliot2003necessary, pettersson2001stabilization} or some known constraints on switching sequence such as dwell time \cite{morse1996supervisory, allerhand2010robust} or average dwell time \cite{hespanha1999stability, zhang2010asynchronously} restrictions.

PDEs are accomplished at describing a wide range of phenomena such as fluid dynamics and quantum mechanics that cannot be adequately described by ODEs \cite{evans2010partial, bao2021partial, bao2022physics}. As an example, a typical application for PDEs is the modelling of congestion in highway networks  \cite{bayen2004network}, and one popular model for highway control is the Lighthill-Whitham-Richards (LWR) model \cite{lighthill1955kinematic,munoz2003traffic}.  Within this context, the PDEs evolve in an infinite dimensional space and must be discretized for computational purposes. 
In our approach, we seek to extend the existing framework from ordinary differential equations to partial differential equations. However, this extension is non-trivial since the internal structure of HA with PDEs is inherently more complicated due to the fact that switching and jump conditions cannot be easily specified. An in-depth discussion of these complexities, along with a discussion of three types of switching conditions, can be found in the following paper by \cite{claudel2008solutions}.

\textcolor{black}{While the hybrid automata theory for ordinary differential equations (ODEs) has been extensively studied (e.g., see \cite{Henzinger1996, Alur1991}), the theory for partial differential equations (PDEs) is still in the stage of case study \cite{bao2023modelling, Mallet2006, Banerjee2013, Tomlin1999}. The main challenges in this area can be attributed to two factors:
\begin{enumerate}
\item[(1)] The dynamics of automata involve switching due to known constraints, which gives rise to the so-called \emph{free boundary problem}. This problem entails tracking the interface between different stages. In the context of PDEs, analyzing and numerically solving the free boundary is highly challenging (e.g., see \cite{Bonnerot1977, Aitchison1985, Friedman1988, Caffarelli2005, Friedman2000, Ryskin1984, Ryskin1984a}).
\item [(2)] Unlike ODEs, PDEs incorporate various types of boundary conditions. Determining suitable boundary conditions for the partition is a complex task that poses numerical challenges.
\end{enumerate}
Therefore, in this article, we present a novel approach to address the hybrid automata problem with partial differential modelling. Our approach aims to reduce the PDE problem to a manageable ODE hybrid automaton by employing a standard numerical discretization technique known as the \emph{finite difference method}.
}

%% file: PDHA.tex
\section{Running Examples}\label{sec:re}
\begin{figure}[t!]
\centering
    \includegraphics[width=14cm]{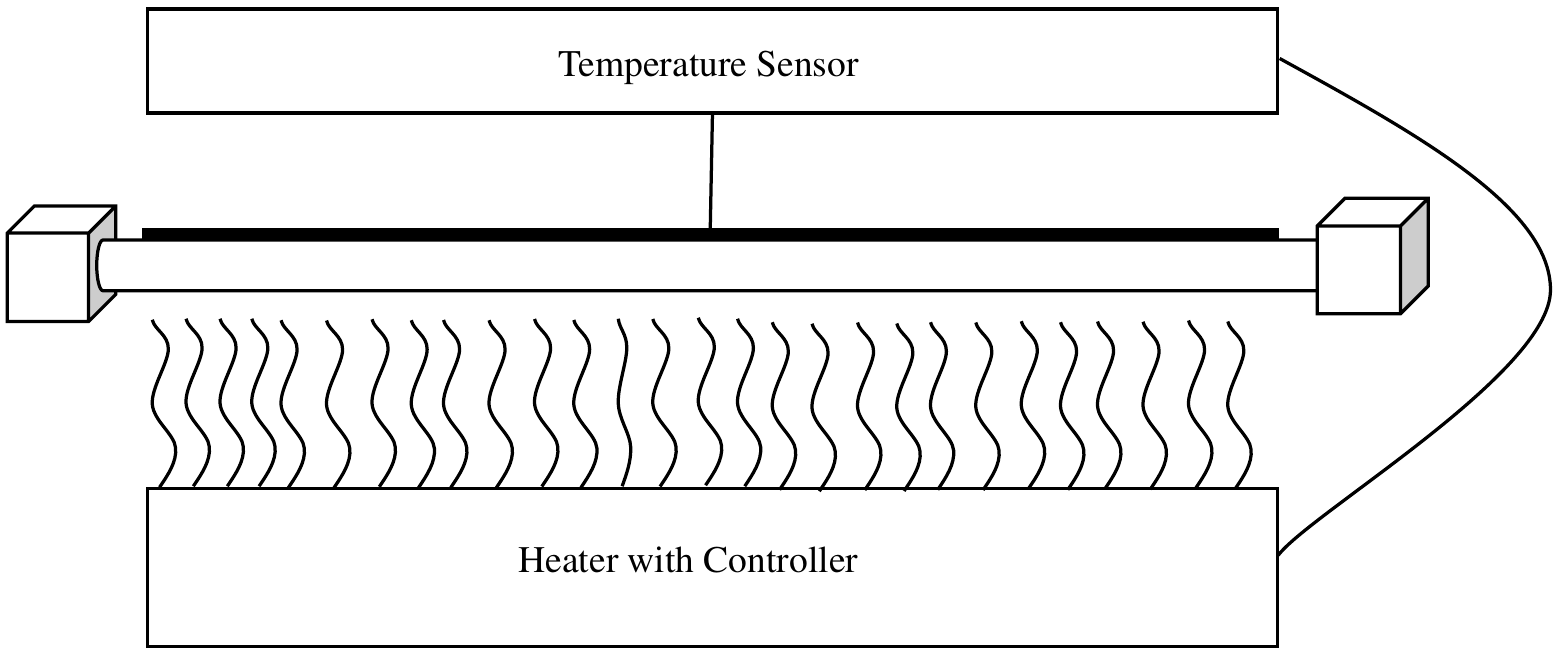}
    \caption{Diagram for heater model.}
    \label{fig:heater}
\end{figure}

\textcolor{black}{Before illustrating the main ideas} of our framework, we use two running examples throughout the paper. The components of these running examples are depicted in Figures \ref{fig:heater} and \ref{fig:car_highway}. The heater model consists of three devices: (i) a rod that can be heated using a gas burner from anywhere, (ii) a long gas burner that can be turned on or off partially in any region, and (iii) a thermometer that monitors the temperature of the rod at every location and periodically issues signals when the temperature of the rod is above or below certain a threshold. The goal is to design a control policy that maintains the temperature of the rod within a given range.

The second model we consider is a traffic model that was originally proposed in \cite{claudel2008solutions}. The system consists of a highway segment in which vehicles flow in and out. Vehicles maintain a high speed when the traffic flow density is below a given critical density and when the density exceeds the critical threshold, the speed of the traffic wave and the overall flow capacity decrease.

\subsection{Heater Model}

We first describe the behavior of the temperature of the rod. When the gas burner is OFF, the temperature of the rod, denoted by the variable $u$, decreases according to the following PDE: $u(x, t)_t - \alpha u(x, t)_{xx} = 0$ where $\alpha$ is the thermal diffusivity. The thermal diffusivity measures the rate of heat transfer of a material from the hot side to the cold side. The boundary conditions are $u(0, t) = 0$ and $u(L, t) = 0$, which denote that at the location $x = 0$ and $x = L$ the temperature is fixed to be $0$, and heat is absorbed. $X$ denotes the space coordinate and $t$ denotes time. This law, however, is only true when the temperature of the corresponding position on the rod is greater than 0.7. When the heater is OFF, and the temperature of a position on the rod is below 0.4, the heater at that position will be turned on. The heater function is defined as $f(x) = -\frac{x}{L} + 1$. The maximum of the heater function appears at $x = 0$ and the minimum is $0$ at $x = L$. During the heater ON mode, the temperature is governed by the equation $u(x, t)_t - \alpha u(x, t)_{xx} = f(x)$ where $f(x)$ is the function defined above. As one can see from the description of the evolution of the temperature, the system is not purely continuous. The system can switch from one mode to another by turning the heater on and off at a specific location on the rod.

\subsection{Traffic Flow Model}
\begin{figure}[t!]
\centering
    \includegraphics[width = 14cm]{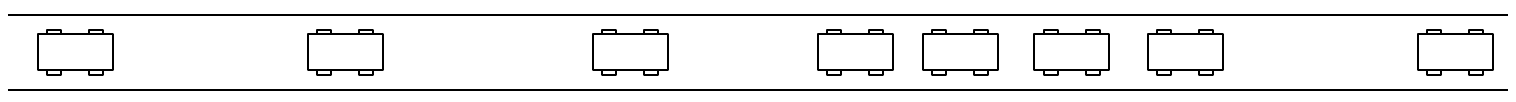}
    \caption{Diagram for traffic model.}
    \label{fig:car_highway}
\end{figure}

The traffic model describes how traffic conditions may behave on a given highway. In free flow mode, which means the density is less than a specified critical density $\rho_c$, the density function is given by the following equation $u(x, t)_t + v_1 u(x, t)_{x} = 0$, where $v_1$ is always a positive number and denotes that the traffic flow propagates forward. However, if the traffic density achieves $\rho_c$, the system transitions into a high-density congestion mode. In this mode, the governing equation is described by $u(x, t)_t + v_2 u(x, t)_{x} = 0$, where $v_2$ is always a negative number and denotes that traffic congestion always moves backwards and propagates upstream. As an example, one of the causes of congestion may be a result of merging slow and fast-moving traffic, resulting in a high-density traffic flow. In this model, we can frequently switch between a free-flowing mode and a congestion mode, and as a result, a more powerful hybrid framework is needed to formalize these kinds of system behavior.

Additionally, the two models we have outlined present a mixture of space and time variables and are therefore quite complicated. In the following sections, we present a framework that is able to capture such a mixture of discrete and continuous behaviors, where PDEs describe the temporal and spatial dynamics.


\section{Partial Differential Hybrid Automata}\label{sec:ha}
\subsection{Transitions, Trajectories and Executions}\label{sec:dt}

\begin{figure}[t!]
\centering
    \includegraphics[width=14cm]{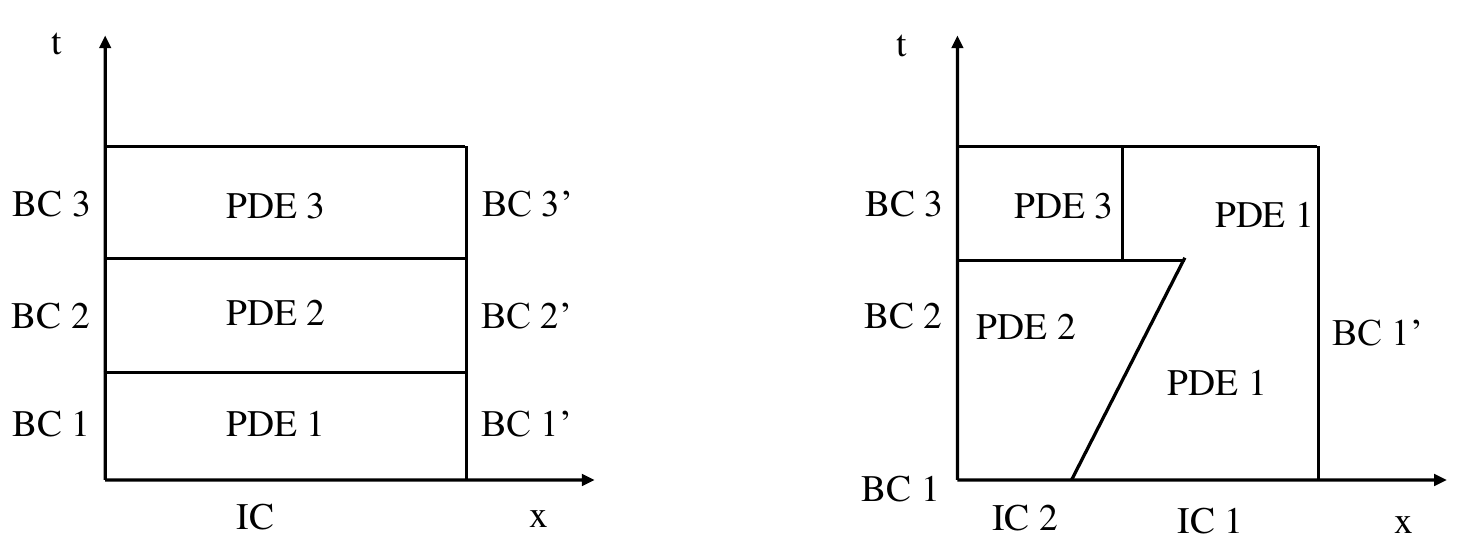}
    \caption{Switching cases for two scenarios, full domain switching on left and partial domain switching on right. BC denotes boundary conditions and IC denotes initial conditions. $x$ axis denotes the spatial domain and $y$ axis denotes time.}
    \label{fig:domainswitch}
\end{figure}

Due to the properties of PDEs, the switching structure is more involved than in the case of ODEs \cite{claudel2008solutions}. One highway congestion alleviation model in \cite{bayen2004network} considers $N$ connected segments, where each segment is governed by an independent LWR PDE \cite{lighthill1955kinematic} with one side boundary condition. Each segment affects the boundary condition of the preceding segment and serves as an input function to the PDE. Additionally, the system is allowed to switch to another mode that incorporates a speed limit. 
One control approach for transport systems \cite{hante2009modeling} proposes rules for switching based on criteria of minimizing a cost function.  
We outline two specific structures that are of interest to our framework:

\begin{itemize}
  \item \emph{Switching PDEs in time on the full spatial domain}. This situation is portrayed in Figure \ref{fig:domainswitch} (left) and is the PDE counterpart of hybrid systems as defined in \cite{tomlin2000game}, where a game theory approach is used to determine when mode switching occurs. Boundary condition switching has been investigated in highway traffic applications in \cite{s2006weak}. PDE switching in time appears in \cite{hante2009modeling} and in \cite{bayen2004network} and the latter one contains switching between modes in each highway segment. 
  \item \emph{Switching PDEs on subsets of the time-space domain}. This is illustrated in Figure \ref{fig:domainswitch} (right). Examples include the LWR PDE, in \cite{bayen2004network} where each highway segment is coupled with a boundary condition at the starting point of itself. 
  This also appears in \cite{lighthill1955kinematic}, where the LWR with triangular flux function can be decomposed into two modes (two one-dimensional wave equations) resulting in a partition of the $(x, t)$ space in regions with forward travelling waves, and regions with backward travelling waves. 
\end{itemize}

In order to describe the partial domain switching case in Figure \ref{fig:domainswitch} (right), it is useful to partition the whole domain into several parts and assign each one a corresponding mode. One possible solution is to use the union of disjoint domains to represent the region governed by specific modes. For example, the heat equation $u_t - u_{xx} = 0$ with $\alpha = 1$ (mode 1) and without heat source can cover domain $[0,0.4)$. The equation $u_t - u_{xx} = f(x)$ with external source (mode 2) occupies domain $[0.4, 1]$.

Based on this idea, the main challenge is how to formalize the particular structure of the domain occupying behavior. One approach is to associate the union operator with the modes of the system. Thus, the system mode $q$ can be described as a composition of multiple modes $\cup q_i$, where each mode specifies the domain it occupies. Like in $q_1$, $u_t - u_{xx} = 0$ when $x \in \mathcal{D}_1$ and in $q_2$, $u_t - u_{xx} = f(x)$ when $x \in \mathcal{D}_2$, where $\mathcal{D}_i$ is the region occupied by $q_i$. Thus, the space domain $X$ is composed of $\mathcal{D}_1$ and $\mathcal{D}_2$.

This embodies our approach to represent the multiple mode occupation phenomenon, and we refer to the multiple mode occupation as the changing behavior of $\cup \mathcal{D}_i$. The remaining challenge is to how to describe the moving behavior of the modes. This can be realized by changing $\mathcal{D}_i$, together with its boundary condition. As the system evolves, points satisfying a given condition will be moved from one mode to another and cause $\mathcal{D}_i$ to grow or shrink. We call this point's moving behavior a \emph{trajectory}. The evolution of $\mathcal{D}_i$ is not easily depicted. It can be a function of time or a function of several other complicated variables. Additionally, it might be impossible to predict switching behavior due to the complexity exhibited by the PDE's solution.

Based on the ideas discussed above, the concept of \emph{spatial control mode} is given below.

\begin{definition}
    [Spatial Control Mode] Let $X\subseteq \mathbb{R}^{n}$ be a space domain. Let $Q$ be a finite set of $m$ control modes, where each mode is denoted as $q_i$. A space control mode consists of a union of $k ~ (k \leq m)$ control modes, where each mode covers part of the space domain $X$. The spatial control mode is denoted by $\tilde{q}$ where $\tilde{q} = \cup_{q_i \in Q} \{q_i\} \subseteq Q$ and the region mode $q_i$ covers is denoted by $Dom(q_i)$ and $Dom(\cdot)$ is a set map $Dom(\cdot) : Q \to 2^X$. Additionally, $\cup_{i = 1}^{k} \mathcal{D}_i = X$, where $\mathcal{D}_i = Dom(q_i)$. 

\end{definition} 

Now, according to our definition, $\tilde{q}$ is a subset of $Q$ and not just an element in $Q$. We want to emphasize that the system could occupy several control modes simultaneously, rather than stay at a single mode as defined in classical hybrid automata. In order to accommodate $\tilde{q}$ and $Dom(q_i)$ and for the purposes of formalizing domain division, we rely on \emph{partition} defined in Definition \ref{ppt}. One can see that $\cup \{ \mathcal{D}_i\}$ is a \emph{partition} of domain $X$ since each $\mathcal{D}_i$ is nonempty, $\bigcup \{ \mathcal{D}_i\} = X$ and $\mathcal{D}_i \cap \mathcal{D}_j = \emptyset$ for $i \neq j$.  For convenience, we denote $p = \cup \{ \mathcal{D}_i \}$ a specific \emph{partition} of $X$ consisting of $\mathcal{D}_i$s and denote $P = \cup \{ p \}$ a set containing all possible \emph{partition}s of domain $X$. 

\begin{definition}[Partition \cite{halmos2017naive}] A partition of a set $S$ is a set of nonempty subsets of $S$ such that every element $x$ in $X$ is in exactly one of these subsets. 
Equivalently, a family of sets $p$ is a partition of $X$ if and only if all the following conditions hold:
\begin{itemize}
    \item The family $p$ does not contain the empty set (that is $\emptyset \notin p$).
    \item The union of the sets in $p$ is equal to $X$ (that is $\cup_{a \in p} a = p$).
    \item The intersection of any two distinct sets in $p$ is empty (that is ($\forall a,b \in p$) $a \neq b \Rightarrow a \cap b = \emptyset$).
\end{itemize}\label{ppt}
\end{definition} 

It is worth noting that the new PDE dynamic hybrid automaton needs a function to describe its state value. The additional spatial variable requires us to specify the values at different space locations while, in classical hybrid automata, the system only has finite values since the system only considers finite continuous state variables. The values of these state variables and discrete modes can be determined as long as the time is fixed, and the system is deterministic. 
\textcolor{black}{In our model, for any time $t$, we regard that the PDE solution $u(\cdot, t)$ lies in the state space $U=\mathbb{R}^X$. In other words, for any given $x\in X$, $u(x,\cdot)$ represents a real-valued trajectory. }

After we obtain $\tilde{q}$, $p$ and $u(\cdot, t)$, the \emph{state} can be defined now.

\begin{definition}
[State] A state $s$ is a collection of variables that describes the current condition of a PDE dynamic system. It consists of a \emph{space control mode} $\tilde{q}$, a \emph{partition} $p$ of a domain $X$ and a real function $u(\cdot, t)$ defined on $X$.
    \begin{equation}\label{eq:exp}
    \begin{split}
    &s = \tilde{q} \times  p \times u(\cdot, t) \in 2^Q \times P \times U\\
    \end{split}
    \end{equation}

\end{definition}

\noindent In Figure \ref{fig:domainswitch} (left), each mode occupies the whole space domain $X$ and the system stays in only one mode. We have $q = \{q_i\}, \mathcal{D}_i = X$, $i = 1,2,3$. In Figure \ref{fig:domainswitch} (right), PDE 1 (mode 1) and PDE 2 (modes 2) occupy the whole domain $X$, $q = \{q_1 \cup q_2\}$ and $\mathcal{D}_1 \cup \mathcal{D}_2= X$ at $t = 0$. It is worth stressing the differences between representing a state in PDE dynamic hybrid automata and classical hybrid automata. For discrete modes, the combination of $\tilde{q}$ and $p$ replaces the single $q$ and describes the space occupation behavior.

Bearing the above in mind, the next step is to formalize the flow function.

\begin{definition}
[Flow] A flow for a PDE dynamic hybrid automaton is of the form $2^{Q} \times P \times U \rightarrow  U$ which maps from one state to another as time evolves. 
\end{definition} 

Similar to classical hybrid automata,  PDE equations simulate the dynamics of the system. The flow consists of a piecewise PDE equation and the corresponding domain partition. A simple flow function for the heater example can be formally represented as

\[
  \begin{cases}
    u_t - u_{xx}= f(x, t; q),~ &0 < x < L,\quad 0<t<T,\quad q\in Q,\\
    u(0, t) = u(L, t)= 0, &0\le t\le T, \\
    u(x, 0)=u_0(x), & 0\le x\le L. 
  \end{cases}
\]

\begin{definition}
[Transition] A transition in a PDE dynamic hybrid automaton is of the form $2^{Q} \times P \times U \times E \rightarrow 2^{Q} \times P$ which maps the current mode and partition to another. $E$ is a finite set of events.
\end{definition}

\begin{figure}[t!]
\centering
    \includegraphics[width=14cm]{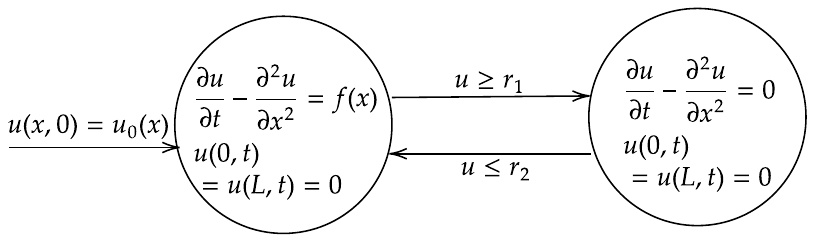}
    \caption{Heater PDHA corresponding to the system in Figure \ref{fig:heater}.}
    \label{fig:automata1}
\end{figure}

\begin{definition}
	[Reset] A reset function for a PDE dynamic hybrid automaton is of the form $2^{Q} \times P \times U \times E \rightarrow U$ which specifies how the value of $u(x, t)$ changes when a transition takes place.
\end{definition}
\textcolor{black}{ The ON mode, triggered by conditions defined in the set $E$, for example (Figure \ref{fig:automata1}), $u(x, t)\ge r_1$, then $q=q_1$ so that $f(x, t; q_1)=f(x)$ while the OFF mode happens when $u(x, t)\le r_2$, we set $q=q_2$, then $f(x, t; q_2)=0$. }






\subsection{Partial Differential Hybrid Automaton (PDHA)}\label{sec:ss}

With the above discussion in mind, we can now provide a formal definition of a partial differential hybrid automaton.

\noindent
\begin{definition}
[Partial Differential Hybrid Automaton (\text{PDHA})] A \emph{partial differential hybrid automaton} is a tuple $\langle Q, E, P, X, U, Init, Inv, f, \phi, G, R \rangle$ where: 
\begin{itemize}
  \item $Q$ is a set of finite $q_i$s that represents control modes.
  \item $E$ is a finite set of events.
  \item $X$ is a space domain and $X \subseteq \mathbb{R}^n$. 
  \item $P$ is a set of all partitions defined on domain $X$.  
  \item $U$ is a function space defined on domain $X$ and $U = \mathbb{R}^{X}$.
  \item $Init$ is a set of initial states and $Init \subseteq 2^Q \times P \times U$.
  \item $Inv$ is a set of invariants defined for each mode $q_i$.
  \item $f : 2^Q \times P \times U \to U$ is a flow function and $f$ has the form :
\[
  \begin{cases}
  u_t = f_1(\Delta u, \nabla u, x, t), ~~ x \in \mathcal{D}_1,~~ q=q_1\\
  u_t = f_2(\Delta u, \nabla u, x, t), ~~ x \in \mathcal{D}_2,~~q=q_2\\
  ...\\
  u_t = f_k(\Delta u, \nabla u, x, t), ~~ x \in \mathcal{D}_k, ~~ q=q_k.\\
  \end{cases}
\]

  where $f_i, i \in \{1,2,...,k\}$, is the continuous segment in $f$, $\Delta$ is the Laplacian, $\nabla$ is the gradient, and $\mathcal{D}_i, i \in \{1,2,...,k\}$ is the domain covered by $q_i$.
  \item $\phi: 2^{Q} \times P \times U \times E \rightarrow 2^{Q} \times P$ is a transition map. 
  \item $G$ is a set defining guard condition, $G \subseteq 2^{Q} \times P \times 2^{Q} \times P \times U$.
  \item $R: 2^{Q} \times P \times U \times E \rightarrow U$ is a reset function.

\end{itemize}
\end{definition}

In our framework, each mode contains only one PDE instead of a system of equations as in classical hybrid automata. Boundary conditions must not be violated and can be treated as a kind of \emph{invariant}. They are imposed on the boundary of $X$ and act like constraints within the mode. Additionally, they are forced to be satisfied even during mode discrete transitions.

\begin{figure}[t!]
\centering
    \includegraphics[width=14cm]{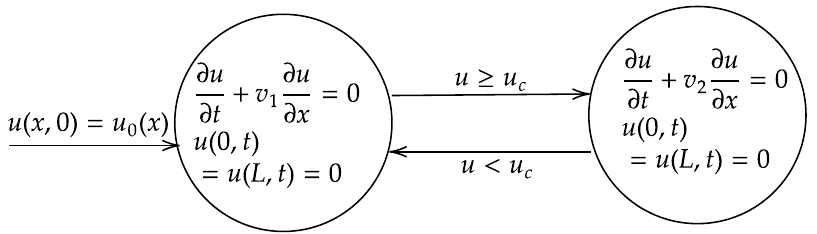}
    \caption{Traffic PDHA corresponding to the system in Figure \ref{fig:car_highway}.}
    \label{fig:automata2}
\end{figure}%


Revisiting the examples outlined at the very beginning, we construct two models to illustrate the idea of PDHA in  Figures \ref{fig:heater} and \ref{fig:car_highway}. In each model, the system is composed of two modes and each mode is governed by a PDE, the given boundary conditions, and the initial condition. In the heater model, we have $f_i(\Delta u, \nabla u, x, t)=\Delta u+f(x, t; q_i)$. 

The continuous trajectory is realized by moving a point satisfying a guard condition $G$ to another mode. In the heater model, a point in the mode ON where the temperature exceeds the threshold will be transferred to the mode OFF, meaning the heater located at this point is turned off while the other areas remain unchanged. At the same time, as the temperature at some points drops below some given bound, the heater at that location is turned on and the point at mode OFF will be moved to mode ON. 

The same rule applies to the traffic model as well. The points in the free flow mode satisfying the guard condition $G_1$ will be moved to a congestion mode, and points in the congestion mode satisfying $G_2$ will be moved to the free flow mode.

%% file: DSPDHA.tex
\section{Discrete Space Partial Differential Hybrid Automaton}

In this section, we describe some core steps in deriving DSPDHA. 
First, we define a discretization scheme and use it to discretize a PDHA to construct its ``discrete'' model for analysis. Second, using this discretization scheme, we define a discrete PDHA that is related to the original PDHA via a ``discretization relation'' corresponding to the scheme $\mathcal{R}$.

\subsection{Discretization Scheme and Discretization Relation}

A discrete treatment of PDEs requires the use of new tools to realize the functions we need. One of the most widely used sets of methods is called finite difference methods (FDMs). Generally speaking, FDMs are a family of numerical approaches for solving differential equations by replacing them with an approximation using difference equations. Combining FDMs and PDHA produces the following definition:

\begin{definition}[Discretization Scheme]
A discretization scheme is a numerical scheme to discretize a continuous partial differential equation into a set of algebraic equations (full-discretization scheme) or ordinary differential equations (semi-discretization scheme).

\end{definition}

We present several schemes below to demonstrate how this technique approximates spatial derivatives.

\begin{example}[Finite Difference Method Discretization Scheme]
A common example is shown for the heat equation using the central difference scheme:
\begin{equation}\label{eq:cts}
\begin{split}
u(i \times h,t)_{xx} = \frac{u((i - 1)\times h,t) - 2u(i \times h, t) + u((i + 1)\times h, t)}{{h}^2} + O(h^2), 
\end{split}
\end{equation}
\end{example}
\noindent where the scheme is based on a uniform mesh,  $h$ is the grid size, $h = x_{i + 1} - x_i$, $\{x_1, x_2, x_3,..., x_n \}$ is a list of locations where the mesh points sit, and these points are the same ones defined in $p_d$. Additionally, $u(i \times h, t)$ denotes the value at the $i^{th}$ position given time $t$ and the respective derivative is approximated using itself and the two adjacent values. The proof is based on using Taylor's expansion with error $O(h^2)$.

After applying (\ref{eq:cts}) and dropping the redundant part $O(h^2)$, the heat equation becomes
\begin{equation}\label{eq:cts2}
\begin{split}
&\dot{u_{i}}(t) \approx \frac{1}{{h}^2}(u_{i - 1}(t) - 2u_{i}(t) + u_{i + 1}(t)) + f(i \times h,t),~~ i=1, 2, \cdots, n. 
\end{split}
\end{equation}

We replace $u(i \times h, t)$ by $u_i(t)$ since the derivative with respect to space has been removed. Thus, $\dot{u}_{i}(t)$ is used to represent the time derivative at the grid point $i \times h$, and $f(i \times h, t)$ represents the input function valued at the same location. 


Once the scheme has been chosen, the relation between our models can be described by the following definition.

\begin{definition}[Discretization Relation]
If the model $A$ is obtained from the model $B$ via discretization scheme $\mathcal{R}$, we say that the model $A$ relates to the model $B$ via discretization scheme $\mathcal{R}$, which is denoted by: $A \preceq^{\mathcal{R}} B$.  
\end{definition}

\subsection{Discrete Space Partial Differential Hybrid Automaton}

Using a discretization scheme, we obtain discrete models of the state, transition, flow and reset function of the original PHDA. These discrete models are defined below. 

\begin{figure}[t!]
\centering
    \includegraphics[width = 15cm]{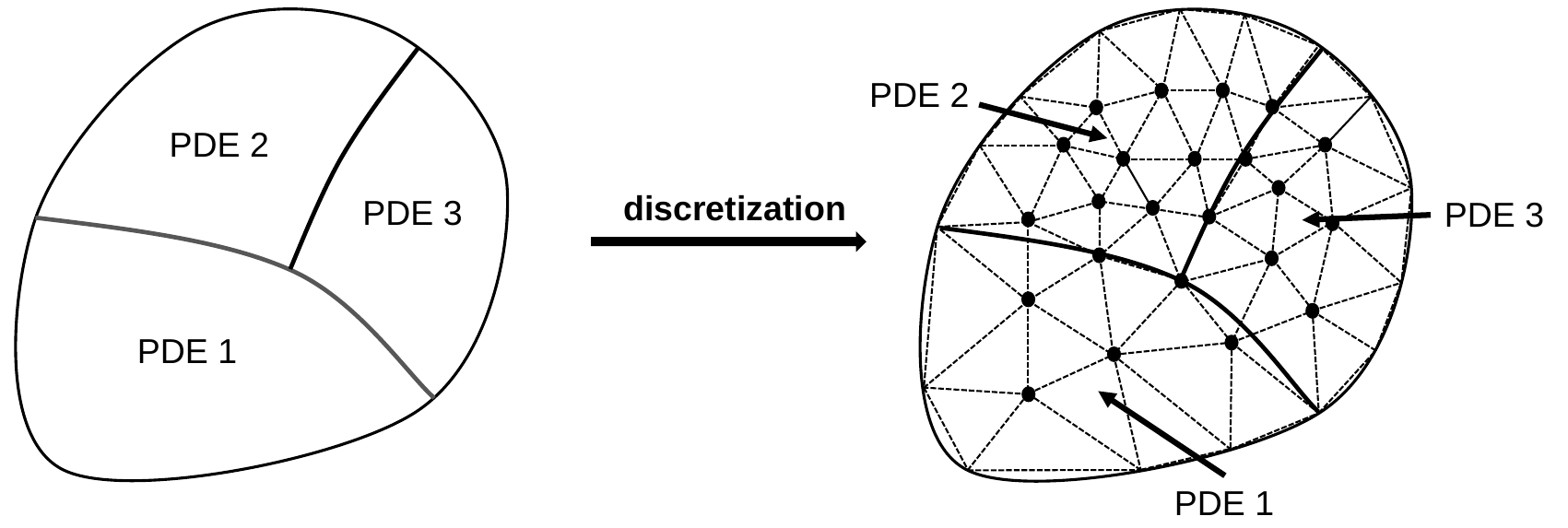}
    \caption{Irregular discretization on domain $X$, the left diagram shows the partition of $X$ and the right diagram shows how discretization is used on the domain $X$. The black dots represent the points chosen in $X_d$ and are connected through dash lines. These points can be on the boundary or inside the region.}%
    \label{fig:irregu_mesh}%
\end{figure}


\begin{definition}
[Discrete Domain] Let $X$ be a domain and $X \subseteq \mathbb{R}^{n}$, a discrete domain $X_d$ of $X$ is a set of distinct points obtained from $X$ by a discretization scheme $\mathcal{\mathcal{R}}$: $X_d = \bigcup_{i = 1}^{m} \{x_i\}$, $x_i \in X$ and $m$ is the number of points. We say $X_d \preceq^{\mathcal{R}} X$.
\end{definition}

\begin{example}[Discrete Domain of Heater Model]
Consider the heat equation defined on domain $X = [0, 10]$. One possible corresponding discrete domain $X_d$ is $\{0, 1, 2, 3, 4, 5, 6, 7, 8, 9, 10\}$.
\end{example}

\begin{definition}
[Discrete Partition] Let $Q$ be a set of modes, $P$ is the set of all partitions on $X$ and $X_d \preceq^{\mathcal{R}} X$. A discrete partition $p_d \in Q^{m}$ is a set of values, and each value is associated with a point $x_i \in X_d$. The set of all possible $p_d$s is denoted by $P_d = Q^{m}$ and $P_d \preceq^{\mathcal{R}} P$.
\end{definition}

In contrast to the continuous domain hybrid automaton which uses $\tilde{q} \times p$ to describe the current mode that the system occupies, a discrete partition combines the two notations and simplifies the expression.

\begin{example}[Discrete Partition of Heater Model]
 Consider the model with $\ q_1 =$ OFF, $\ q_2 = $ ON and $X_d = \{ 0, 1, 2, 3, 4, 5 \}$. Then $\tilde{q} = \{q_1, q_2\}$, $\mathcal{D}_1 = [0,3]$ and $\mathcal{D}_2 = (3,5]$ can be expressed as $\{q_1, q_1, q_1, q_1, q_2, q_2\}$.
\end{example}

\begin{definition}
[Discrete State] A discrete state $s_d$ is an ordered pair $(p_d, u_d)$ where $p_d \in P_d \preceq^{\mathcal{R}} P$ is a discrete partition on $X_d \preceq^{\mathcal{R}} X$ and $u_d \in \mathbb{R}^m$ is a list of real values. We denote the set of all possible $s_d$s by $S_d$ and say $S_d \preceq^{\mathcal{R}} S$ where $S$ is the set of states defined in PDHA.
\end{definition}

A discrete state gives the whole picture of the system by combining a discrete partition and state values.

\begin{example} [Discrete State of Heater Model]
One discrete state can be in the form $\{ \{q_1, q_1, q_1, q_1, q_2, q_2\}, \\ \{0.6, 0.6, 0.5, 0.6, 0.7, 0.8\} \}$, where $q_1, q_2$ are the modes and the succeeding numbers are the temperatures at each node.
\end{example}

Everything listed above leads us to \emph{discrete flow}.

\begin{definition}
[Discrete Flow] Let $f$ be a flow function, $f_d$ is the set of discrete flow functions such that the spatial derivative of $f$ is approximated using $\mathcal{R}$. We say $f_d \preceq^{\mathcal{R}} f$.
\label{def:dicrt_flow}
\end{definition}

\begin{example}[Discrete Flow of Heater Model]
\textcolor{black}{Consider the heater example again, 
\[
  \begin{cases}
    u_t - u_{xx}= f(x, t),\quad &0<t<T,\quad  0 < x < L, \\
     u(0, t) = u(L, t)= 0, \quad &0\le t\le T.
  \end{cases}
\]}
The spatial derivative $u_{xx}$ is approximated by a FDM centred difference scheme based on a uniform mesh \textcolor{black}{of step size} \textcolor{black}{$\Delta x$}. \textcolor{black}{We obtain a system of ODEs:} \textcolor{black}{In place of $u(t, i\Delta x)$, we obtain an ODE system for $u_i(t)$, $i=1, \cdots, m$, $(m+1)\Delta x=L$:}

\textcolor{black}{
$$\dot{u}_i(t)=\frac{u_{i-1}-2u_i+u_{i+1}}{\Delta x^2}+f(t, i\Delta x),$$
where $u_0=u_1, u_{m+1}=u_m$ due to the boundary condition. \textcolor{black}{We would like to remark that the convergence of FDM is due to the Lax--Richtmyer theorem \cite{Smith1985}.}
}
\end{example}


\begin{definition}
[Discrete Transition] A discrete transition $\phi_d$ of a DSPDHA is of the form $P_d \times U_d \times E \rightarrow P_d$ which maps a current discrete partition to another. We say $\phi_d \preceq^{\mathcal{R}} \phi$ where $\phi$ is the transition defined in the original $PDHA$ and $\mathcal{R}$ is the discretization scheme.
\end{definition}

\begin{example}[Discrete Transition of heater model]

In order to illustrate the idea of a discrete transition, we make a simple modification of the previous example:

\[
      \begin{cases}
          \dot{u}_1 =\displaystyle \frac{-2u_1+u_2}{h^2}, \\
          \dot{u}_2 =\displaystyle \frac{u_1-2u_2+u_3}{h^2}, \\
         \quad \vdots\\
      \displaystyle    \dot{u}_m = \frac{u_{m-1}-2u_m}{h^2}. 
      \end{cases}
\]

\[
      \xrightarrow{e^{'}}
      \begin{cases}
          \dot{u}_1 = \displaystyle \frac{-2u_1+u_2}{h^2}, ~~ \\
          \dot{u}_2 = \displaystyle \frac{u_1-2u_2+u_3}{h^2}+f(2h, t), ~~ \\
          \quad\vdots\\
       \displaystyle   \dot{u}_m = \frac{u_{m-1}-2u_m}{h^2}+f(mh, t).  
      \end{cases}      
\]

A transition to the mode ON at location $x_1$ and $x_2$ occurs because event $e$ takes place. Thus, we can write the transition function
\begin{equation}
    \phi_d(p_{d,1}; u_d; e) = \begin{cases}
          p_{d,2},~~ e = e^{'},\\
          p_{d,1},~~ otherwise.\\
      \end{cases} 
\end{equation}
where $p_{d,1} = \{q_1,q_1,q_2,...,q_2\}$, $p_{d,2} = \{q_2,q_2,q_2,...,q_2\}$, $q_1$ represents mode OFF and $q_2$ represents mode ON.  
\end{example}

\begin{definition}
[Discrete Reset] A discrete reset function $R_d$ of a DSPDHA is of the form $P_d \times U_d \times E \rightarrow U_d$ which specifies how a value of $u_d$ changes to a new value when a transition takes place. We say $R_d \preceq^{\mathcal{R}} R$ where $R$ on the right is the reset function in original $PDHA$.
\end{definition}

\begin{definition}
[Discrete Space Partial Differential Hybrid Automaton (\text{DSPDHA})] A \emph{discrete space partial differential hybrid automaton} is a tuple $\langle Q, E, P_d, X_d, U, Init,  Inv, f_d, \phi_d, G, R_d \rangle$ where: 
\label{dspdha}

\begin{itemize}
  \item $Q$ is a set of finite modes;
  \item $E$ is a set of finite events;
  \item $X_d$ is a set of $m$ points;
  \item $P_d$ is a set of values defined on domain $X_d$, $P_d \in Q^{m}$;  
  \item $U_d$ is a set of discrete values defined on $X_d$, $U_d = \mathbb{R}^{m}$;
  \item $Init$ is a set of initial states, $Init \subseteq P_d \times U_d$;
  \item $Inv$ is a set of invariants defined for each mode $q_i \in Q$;
  \item $f_d : P_d \times U_d \to U_d$ is a set of discrete flow functions and $f_d$ have the form :
    \begin{equation}\label{eq:sys_odes}
      \begin{cases}
          \dot{u}_1 = f_1(u_1, u_2,...,u_m, t), \\
          \dot{u}_2 = f_2(u_1, u_2,...,u_m, t), \\
          ...\\
          \dot{u}_m = f_m(u_1, u_2,...,u_m, t).\\
      \end{cases}
    \end{equation}  
where $\dot{u}_i$ denotes time derivative of $u_i$;

  \item $\phi_d: P_d \times U_d \times E \rightarrow P_d$ is a transition function; 
  \item $G$ is a set defining guard condition, $G \subseteq P_d \times P_d \times U_d$;
  \item $R_d: P_d \times U_d \times E \rightarrow U_d$ is a reset function.
\end{itemize}
\end{definition} 

A key observation is $DPH \preceq^{\mathcal{R}} PH $ if the discrete space partial differential hybrid automaton $DPH$ is obtained from a partial differential hybrid automaton $PH$ through $\mathcal{R}$. 

The DSPHDA provides us with a set of properties that can easily be formalized and stated. Additionally, similar concepts that appear in HA are outlined below.

\begin{definition}
[Hybrid Time Trajectory] A hybrid time trajectory $\tau = \{I_i\}_{i=0}^{N}$ is a finite or infinite sequence of intervals of the real line, such that
\begin{itemize}
    \item $I_i = [\tau_i, \tau_i^{'}]$ for $i < N$;
    \item if $N < \infty$, then either $I_N = [\tau_N, \tau_N^{'}]$, or $I_N = [\tau_N, \tau_N^{'})$;
    \item for all $i$, $\tau_i \leq \tau_i^{'} = \tau_{i+1}$.
\end{itemize}
\end{definition} 
\noindent where $\tau_i^{'}$ are the time when \emph{discrete transition} takes place.

Now that we have defined the syntax of DSPDHA, we define the semantics in terms of \emph{executions}.

\begin{definition}
[Execution] An \emph{execution} of a DSPDHA is a collection $\chi = (\tau, p_d, u_d)$ satisfying 
\begin{itemize}
    \item Initial Condition: $(p_d(\tau_0), u_d(\tau_0)) \in$ Init;
    \item Continuous Evolution: for all $i$ with $\tau_i < \tau_i^{'}$,  $p_d$ is a constant, $u_d$ is a solution to the ODEs (\ref{eq:sys_odes}) over $[\tau_i, \tau_i^{'} ]$ and for all $t \in [\tau_i, \tau_i^{'} )$, $u_d(t) \in I$;
    \item Discrete Evolution: for all $i$, $(p_d(\tau_{i+1}), u_d(\tau_{i + 1})) \in R(p_d(\tau_i^{'}), u_d(\tau_i^{'}))$.    
    
\end{itemize}
\end{definition}

An execution $\chi = (\tau, p_d, u_d)$ is called \emph{finite} if $\chi$ is a finite sequence ending with a closed interval, \emph{infinite} if $\tau$ is either an infinite sequence, or if $\sum_{i = 0}^{\infty} (\tau_i^{'} - \tau_i) = \infty$, and Zeno if it is \emph{infinite} but $\sum_{i = 0}^{\infty} (\tau_i^{'} - \tau_i) < \infty$.

\begin{definition}
[Zeno Execution] An execution $\chi$ is Zeno if
\begin{equation}
    \lim_{i\rightarrow \infty} \tau_i - \tau_0 = \sum_{i = 0}^{\infty} (\tau_i^{'} - \tau_i) = \tau_{\infty} < \infty
\end{equation}
Here $\tau_{\infty}$ is called the Zeno time.
\end{definition}

We use $\mathcal{E}_{DPH}(p_{d,0}, u_{d,0})$ to denote the set of all executions of DSPDHA with initial condition $(p_{d, 0},u_{d,0}) \in Init$, $\mathcal{E}_{DPH}^{*}(p_{d,0}, u_{d,0})$ to denote the set of all finite executions, $\mathcal{E}_{DPH}^{\infty}(p_{d,0}, u_{d,0})$ to denote the set of all infinite executions, and $\mathcal{E}_{DPH}$ to denote the union of $\mathcal{E}_{DPH}(p_{d,0}, u_{d,0})$ over all $(p_{d,0}, u_{d,0}) \in Init$.

\begin{definition}
[Zeno Discrete Space Partial Differential Hybrid Automaton] A discrete space partial differential hybrid automaton DPH is Zeno if there exists $(p_{d,0}, u_{d,0}) \in Init$ such that all executions in $\mathcal{E}_{DPH}^{\infty}(p_{d,0}, u_{d,0})$ are Zeno.

\begin{equation}
    \lim_{i\rightarrow \infty} \tau_i - \tau_0 = \sum_{i = 0}^{\infty} (\tau_i^{'} - \tau_i) = \tau_{\infty} < \infty
\end{equation}
Here $\tau_\infty$ is called the Zeno time.

\end{definition}

The set of states reachable by a DSPDHA, $Reach_{DPH}$, is defined as
\begin{equation}
  \begin{multlined}
    Reach_{DPH} = \{(\hat{p_d}, \hat{u_d}) \in P_d \times U_d: \exists(\tau, p_d, u_d) \in \mathcal{E}_{DPH}^{*}, 
    (p_d(N), u_d(\tau_N^{'})) = (\hat{p_d}, \hat{u_d}) \}.
  \end{multlined}  
\end{equation}
where $\tau = \{[\tau_i, \tau_i^{'}]\}_{i = 0}^{N}$.

\begin{definition}
[Nonblocking] A DSPDHA is called nonblocking if $ \mathcal{E}_{DPH}^{\infty}(p_{d,0}, u_{d,0})$ is nonempty at all $(p_{d,0}, u_{d,0}) \in Init$.
\end{definition}

\begin{definition}
[Deterministic] A DSPDHA is called deterministic if $ \mathcal{E}_{DPH}^{*}(p_{d,0}, u_{d,0})$ contains at most one element for all $(p_{d,0}, u_{d,0}) \in Init$.
\end{definition} 
\textcolor{black}{More definitions about DSPDHA are similar to those in the classical HA theory, and we omit them here for the simplicity of the presentation.}

\subsection{Relation to Classical Hybrid Automaton}\label{sec:relation}

In this section, we explore the relations between classical hybrid automata, DSPDHA and PDHA. The central basis of introducing PDHA is extending HA with ODE dynamics to discretize PDE dynamics. To achieve this goal, some adjustments need to be made to accommodate the new framework within the classical hybrid automata scheme. It is our hope, that after these changes, the properties that hold for HA will hold for PDHA as well. This is the basis for the motivation to add an intermediate level (DSPDHA) lying between PDHA and HA, which captures the idea of PDHA and also stays close to classical HA. Using this framework, we make comparisons and illustrate useful results in the form of theorems.

In PDHA, we use space control modes and partitions $\tilde{q} \times p$ to replace a discrete mode $q$, a state function $\varphi(x)$ to replace state variables $X$. The weakness of this scheme is determining executions and transitions because $\tilde{q} \times p$ evolves continuously. DSPDHA overcomes these shortcomings by introducing a discrete state partition, $p_d$ which functions similarly to $q$ in HA. Along the same lines, a discrete flow function obtained from PDHA by FDM is essentially the same as a HA flow function. This allows us to assert some PDE system properties where we cannot otherwise reason with certainty due to a lack of information. We lay out several points below to illustrate how these connections are built.

\begin{figure}[t!]
\centering
    \includegraphics[width = 14cm]{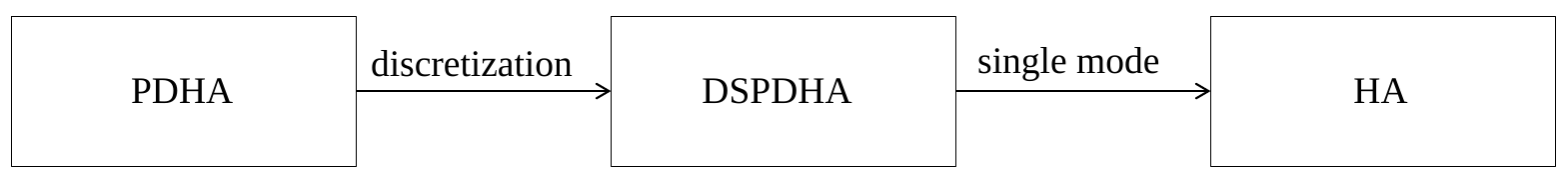}%
    \vspace{-0.5em}%
    \caption{PDHA, DSPDHA and HA relations.\vspace{-0.3em}}%
    \label{fig:relations}%
\end{figure}

\begin{enumerate}
	\item[(1)] \emph{Any DSPDHA is a HA if $p_d$ has identical elements.} Based on the knowledge that elements in $p_d$ represent the modes where each location resides, identical elements indicate all the locations are assigned to the same modes and the system stays at a single mode. Therefore, the given DSPDHA is a HA. \textcolor{black}{Thus, the general dynamical properties of HA, for instance,  the existence, uniqueness, continuous dependence, reachability, stability and invariant sets follows. (See \cite{lygeros2003dynamical}). On the other, the DSPDHA inherits the physical feature (like diffusion) from the PDE model, which we demonstrate in the heat model below. } 
	\item[(2)]\emph{If the order of the time derivative of the PDE considered in a PDHA is $1$, the number of points $m$, in a discrete domain $X_d$ is equivalent to the size of flow functions in the corresponding DSPDHA.} According to Definition \ref{def:dicrt_flow}, the discretization is made on each discrete point in $X_d$. As a result, we have $m$ equations eventually.
	\item[(3)]\emph{If the order of the highest time derivative of a PDE is $n$, then the PDE can be reduced to a system of $n$ PDEs with a $1^{st}$ order time derivative.} We begin the proof by a change of variables. Let us replace the $k^{st}$ order time derivative of $u$ by $v_{k}$ and add a new equation to the system once the replacement is done. In the end, we obtain a system of equations in the form below.
	\begin{equation}\label{eq:order_redc}
		\begin{cases}
			\dot{u} = v_{1}, \\  
			\dot{v}_{1} = v_{2}, \\
			\dot{v}_{2} = v_{3}, \\
			...,\\
			\dot{v}_{n - 2} = v_{n - 1}, \\
			\dot{v}_{n - 1} = f(u, u_x, u_{xx},..., t).
		\end{cases}
	\end{equation}
	where the original equation is $u^{(n)} = f(u, u_x, u_{xx},..., t)$ and $u^{(n)}$ is its $n^{th}$ time derivative. \textcolor{black}{For example, we can deal with the wave equation $u_{tt}-u_{xx}=0$. }
		\item[(4)]\emph{If the order of time derivative of the PDE considered in a PDHA is $n$ and the number of points in discrete domain $X_d$ is $m$, the size of flow functions in the corresponding DSPDHA is $n \cdot m$.} Since $m$ determines the number of mesh points and each point is associated with an ODE ($1^{st}$  order PDE after spatial discretization), $m$ equals the dimension of the system of ODEs which is the flow function.
\end{enumerate}

Besides investigating the relation and distinction between models above, an interesting observation which occurs in classical hybrid automata are discontinuity issues, especially under PDE dynamics. \cite{hante2009modeling} discusses the propagation of discontinuities when the system switches mode. They are numerous reasons why discontinuities occur. For the transport equation with a nonlinear flux function, there does not always exist a smooth solution, and a weak solution has to be constructed to handle a shock or rarefaction wave \cite{thomas2013numerical}. Jump discontinuities for hyperbolic system are discussed in \cite{rauch1981jump}.

The situation discussed in \cite{hante2009modeling} is likely to happen if the discrete transition takes place in our framework. Consider a segment of road governed by a transport equation and the model is $u_t + u_x = 0$. Supposing the switching is triggered, and a new model $u_t + 2u_x = 0$ is introduced, the latter model which involves fast wave speed will push the front flow to move fast, and the traffic will start to accumulate at the boundary of the road which directly leads to a discontinuity.

%% file: example.tex
\section{Experiments on Running Examples}\label{sec:examples}

The heater and traffic examples are formalized as DSPDHA and analyzed in this section. \footnote{Code for the experiments is available at https://drive.google.com/drive/folders/1qnpwlU9WdVNjYm8lsZOjAqFIQhGhH0ME?usp=sharing}

\begin{figure}[t!]
\centering
    \includegraphics[width = 14cm]{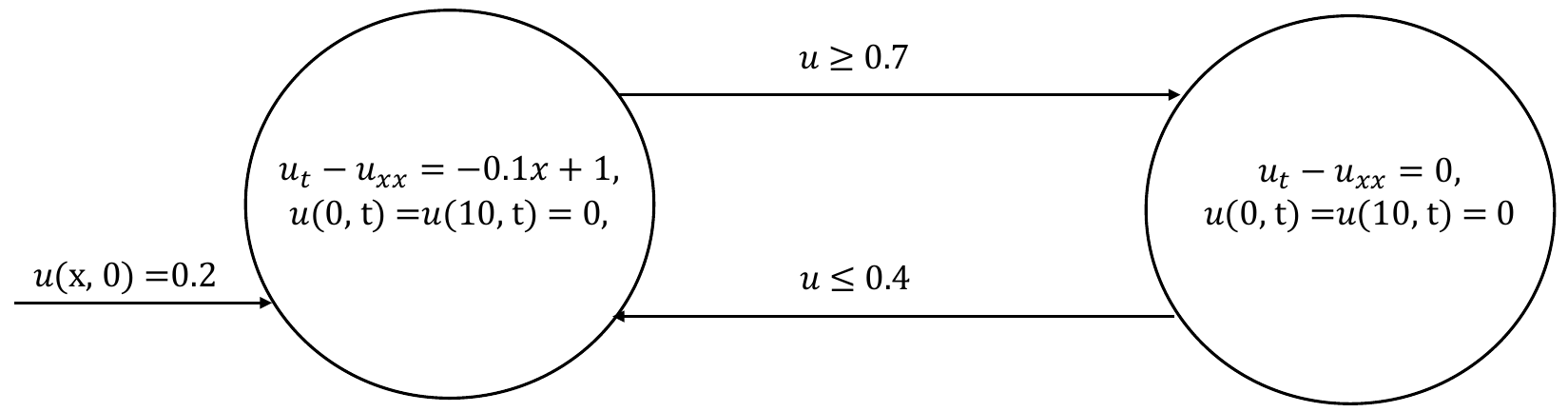}%
    \vspace{-1em}%
    \caption{PDHA diagram for the heater model analyzed. The left circle indicates the ON mode and the right circle indicates the OFF mode. The guard conditions are specified as greater equal than $0.7$ or less equal than $0.4$. The initial condition is $0.2$ for everywhere.}%
    \label{fig:example1}%
\end{figure}

\paragraph*{Heater Model.}%
As illustrated above, the first example is the bronze rod with two isolated ends. The heater provides heat to the rod and can be partially turned on and off at any time once the temperature reaches the given thresholds. For the input function, a linear increment function is considered with a maximum value appearing on the left and a minimum value appearing value on the right. The diagram of the model is in Figure \ref{fig:example1}. 

\begin{figure}
     \centering
     \begin{subfigure}{0.45\textwidth}
         \centering
         \includegraphics[width=6.5cm]{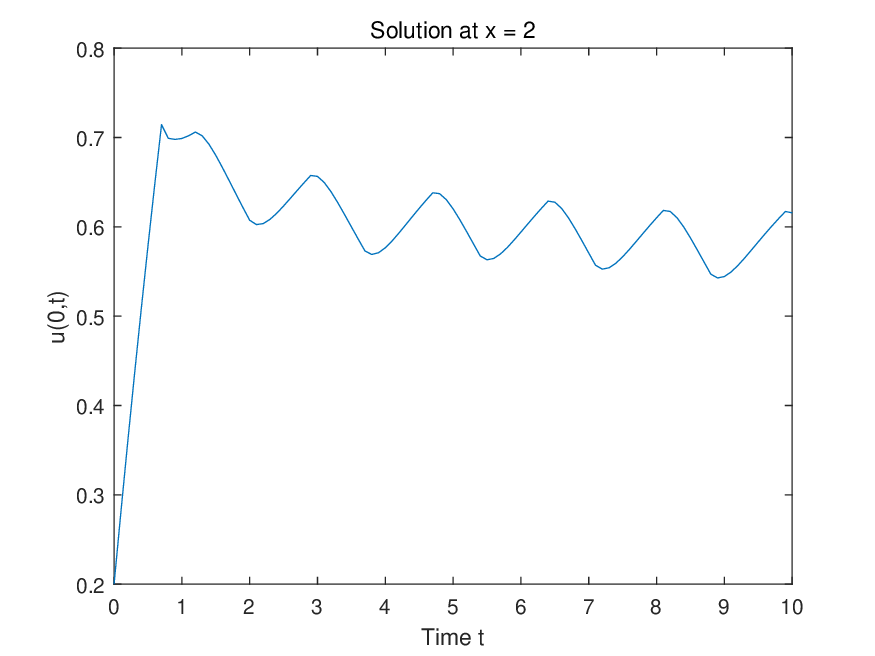}
         \caption{$\Delta x = 1$}
         \label{fig:h1x2}
     \end{subfigure}
     \hfill
     \begin{subfigure}{0.45\textwidth}
         \centering
         \includegraphics[width=6.5cm]{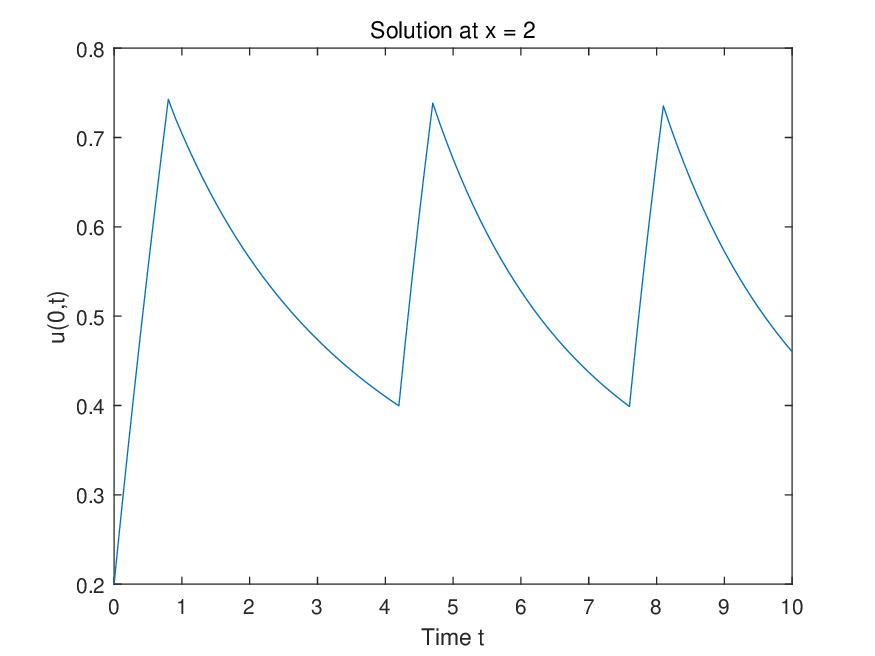}
         \caption{$\Delta x = 2$}
         \label{fig:h2x2}
     \end{subfigure}
     
     \caption{Simulation results of heater model at $x = 2$. The heater at location $x = 2$ is turned on and off repeatedly. The two subfigures use different discretization sizes. }
\end{figure}

\begin{figure}
     \centering
     \begin{subfigure}{0.45\textwidth}
         \centering
         \includegraphics[width=6.5cm]{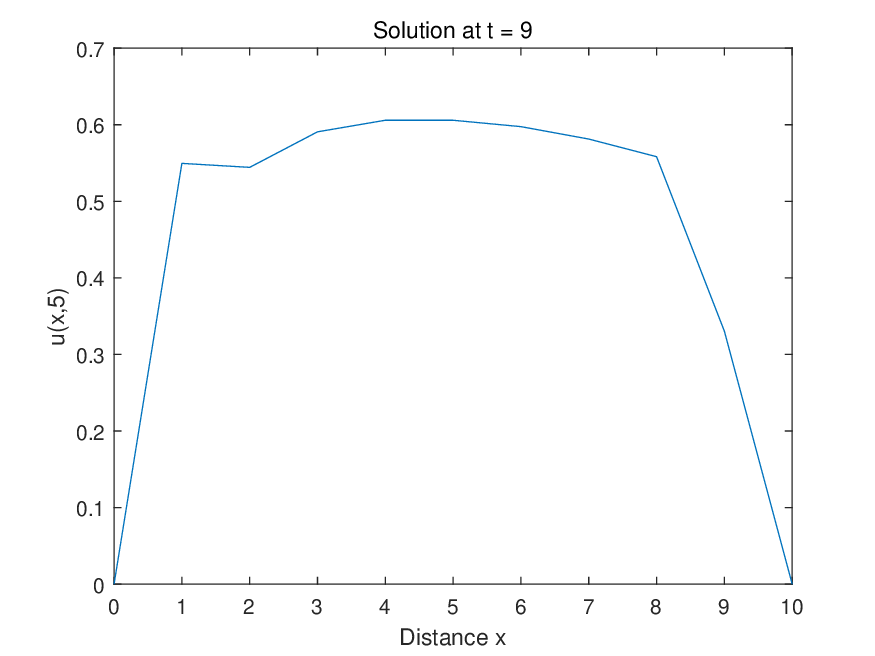}
         \caption{$\Delta x = 1$}
         \label{fig:h1t9}
     \end{subfigure}
     \hfill
     \begin{subfigure}{0.45\textwidth}
         \centering
         \includegraphics[width=6.5cm]{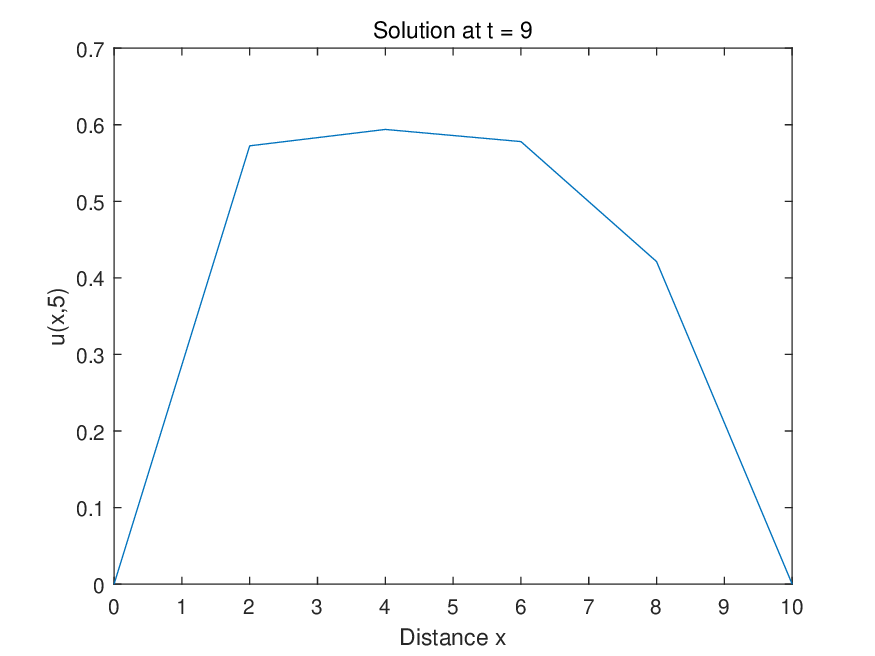}
         \caption{$\Delta x = 2$}
         \label{fig:h2t9}
     \end{subfigure}    
     \caption{Simulation results of heater model at $t = 9$. Temperatures at the boundaries remain $0$. The two subfigures use different discretization sizes. }  
\end{figure}

The idea of \emph{discrete partitions} is illustrated in this example, and $9$ points are chosen to describe the temperature along the rod. A point whose value is greater than 0.7 will be moved to mode OFF and be moved back to ON as the value drops below 0.4. The initial temperature is $0.2$ everywhere on the rod, and the input function for mode ON is $f_1(x) = -0.1x + 1$ and $0$ for mode OFF. Since we have the boundary condition $u(0, t)= 0$ and $u(L, t)= 0$, the two values, $u(x_1, t)$ and $u(x_{n - 1}, t)$, will decrease rapidly due to the heat diffusion effect \cite{evans2010partial}. Additionally, the input function decreases as $x$ increases, therefore the heating effect near the boundary $x = L$ will be trivial, and the heater remains on. We discretize the domain and assign each mesh point a heater. The center difference scheme is used to approximate the space derivative and the forward scheme is used to approximate the time derivative. 

\textcolor{black}{Mathematically speaking, we can see even though the discretization reduces the model in HA for ODE, in the following analysis we can see the dynamic of the solutions is consistent with the diffusion feature of the PDE model: consider the following ODE system
\begin{equation}
  \left\{
  \begin{array}{ll}
   \displaystyle \frac{du_i}{dt}=u_{i-1}-2u_i+u_{i+1}+(10- i) &\text{when }u_i\le 0.4, \\
   \displaystyle \frac{du_i}{dt}=u_{i-1}-2u_i+u_{i+1} &\text{when }u_{i}\ge 0.7, 
  \end{array}
  \right.
  \label{eqn:heatexample}
\end{equation}
where $i=1, 2, \cdots, 9$ and $u_0=u_{10}=0$. Initially, we have $u_{i}(0)=0.2$, \eqref{eqn:heatexample} is an ODE system with constant coefficients, hence the \emph{existence, uniqueness} and \emph{continuous dependence} follows from the standard ODE theory.  Furthermore, it is easy to see that
$$\frac{du_i}{dt}\Big|_{t=0}=-i+10,\quad i=1, 2, \cdots, 9,$$
in other words, $u_i$ is increasing initially since the heat source is turned ON. If there are some $u_i$ hits the threshold value $0.7$, we may pick the maximum value, say $u_i$, then since the heat source is turned OFF, we have
$$\frac{du_i}{dt}=u_{i-1}-2u_i+u_{i+1}\le 0,$$
where we use the fact that $u_{i}\ge u_{i-1}$, and $u_i\ge u_{i+1}$. This implies that $u_{i}$ starts to decrease. For locations that are away from the boundary, if the value falls into $0.4$, we pick the minimum value, then we have that 
$$\frac{du_i}{dt}=(u_{i-1}-2u_i+u_{i+1})+10-i\ge 10-i>0,$$
where we use the fact that $u_i\le u_{i-1}$ and $u_{i}\le u_{i+1}.$ Therefore, $u_i$ starts to increase again. We conclude that \emph{invariant region} for $u_{i}$, $i=2, 3, \cdots, 8$ is $[0.4, 0.7]$. For $u_1$ and $u_9$, we claim that their values cannot fall below $0$, in fact, from the OFF mode, when $u_1$ and $u_9$ go down to 0,  we have
$$\frac{du_1}{dt}=-2u_1+u_2+9=u_2+9>0, \qquad \frac{du_9}{dt}=u_8-2u_9+1=u_8+1>0,$$
which means the values of $u_1$ and $u_9$ will start to go up again which prevents the values fall further into negatives. Hence, the \emph{invariant region} for $u_1$ and $u_9$ is $[0, 0.7]$.  
}

\paragraph*{Traffic Flow Model.}
We consider the model appearing in \cite{claudel2008solutions}. The switched PDE problem is a LWR PDE \cite{lighthill1955kinematic} with the triangular flux function. The flux function and two PDE modes are:%
\begin{equation}\label{eq:traffic_mode}
  \phi(u) = \begin{cases}
    v_1u, &~ x < u_c, \\
    v_2(\omega -u),& ~ x \geq u_c.
  \end{cases}~~
  \text{mode} = \begin{cases}
    u_t + v_1 u_x = 0,~ x < u_c, \\
    u_t - v_2 u_x = 0,~ x \geq u_c.
  \end{cases}
\end{equation}

\begin{figure}[t!]
\centering
    \includegraphics[width=14cm]{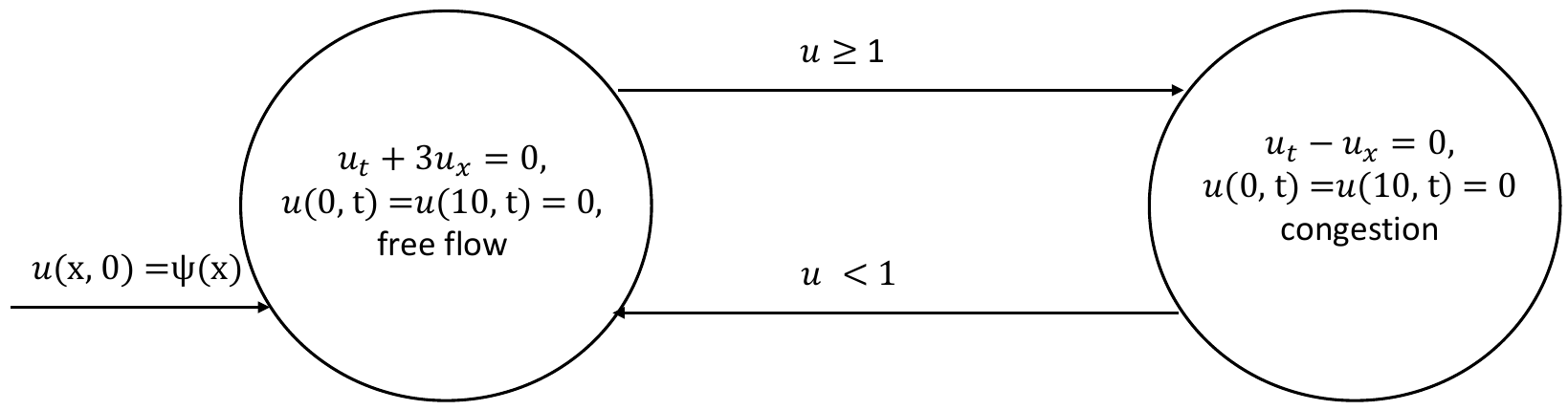}%
    \caption{PDHA diagram for the traffic model analyzed. The left circle indicates the free flow mode and the right circle indicates the congestion mode. The guard conditions are specified as greater equal than $1$ or less than $1$. The initial condition is $\Psi(x)$.}%
    \label{fig:example2}%
\end{figure}

We chose $v_1 = 3$, $v_2 = 1$, $\omega = 4$ and $u_c = 1$. The forward traffic flow maintains a speed equal to $3$ and the backward wave maintains a speed equal to $1$. Moreover, $u_c$ equals $1$ which is the guard condition for switching. Lastly, $\omega$ is the parameter specified in backward wave. The diagram of the model is in Figure \ref{fig:example2}. 

Initially, the regions, $\{0, ...,  0.9\} \cup \{1.3, ...,3.4\} \cup  \{3.8,...,
10.0\}$, are assigned a value $0$ for free flow mode. The other remaining areas, $ \{1.0, 1,1, 1.2\} \cup \{3.5, 3.6, 3.7\}$, are assigned a value $1$ for congestion mode. The Equation \ref{eq:traffic_mode} shows the free flow mode carries the traffic forward and congestion pushes the traffic backwards. Once the forward and backward waves meet each other, the forward wave merges into the backward wave and starts to move back. The traffic leaves the system and no longer enters again when it reaches the boundary. For simplicity, no incoming traffic is considered.

\begin{figure}
     \centering
     \begin{subfigure}{0.45\textwidth}
         \centering
         \includegraphics[width=6.5cm]{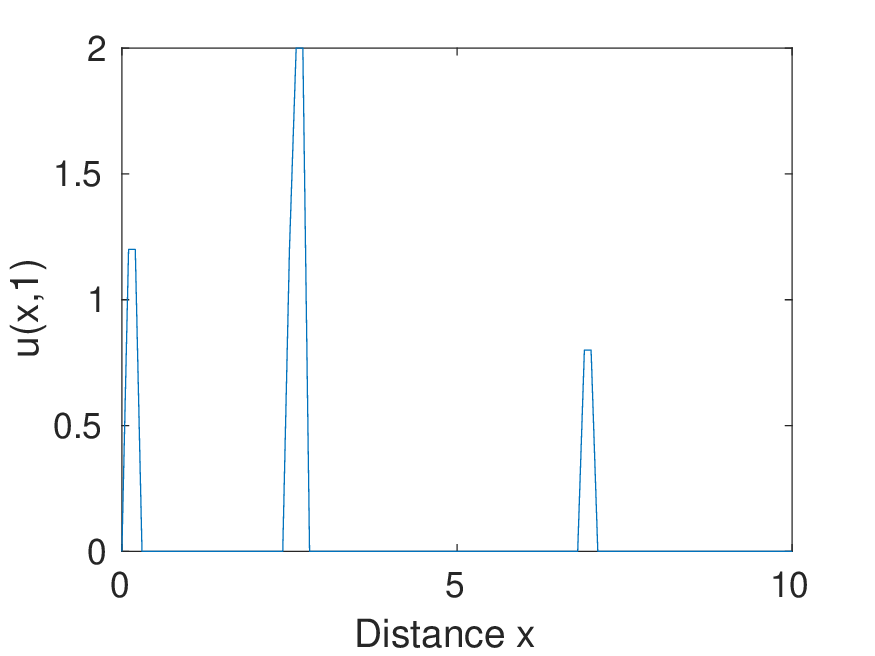}
         \caption{$t = 1$.}
         \label{fig:71}
     \end{subfigure}
     \hfill
     \begin{subfigure}{0.45\textwidth}
         \centering
         \includegraphics[width=6.5cm]{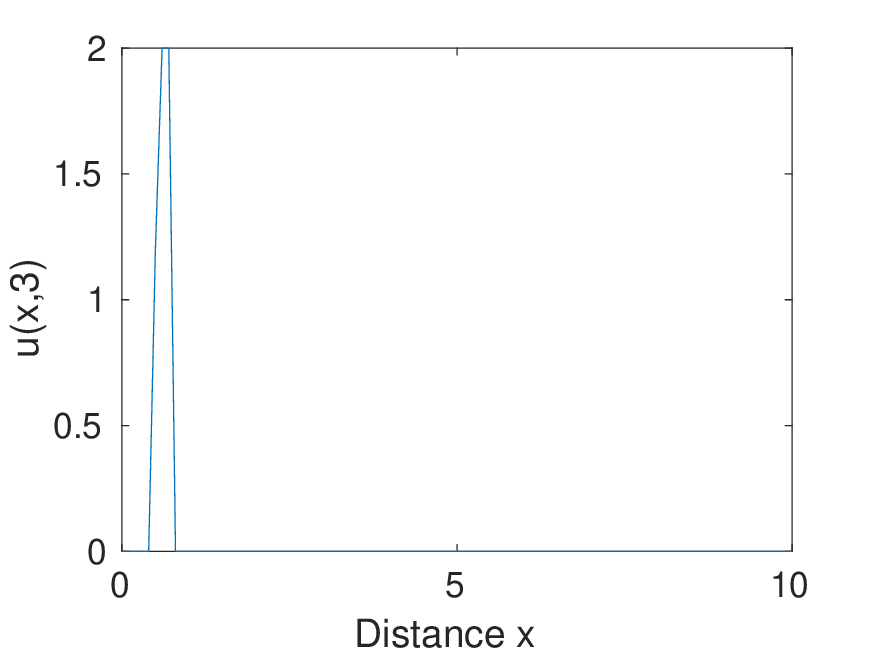}
         \caption{$t = 3$}
         \label{fig:72}
     \end{subfigure}    
     \caption{The left subfigure shows the simulation results of the traffic model at $t = 1$. A backward wave appears at the boundary $x = 0$. A forward wave starting at $[3.9, 4.0]$ has propagated to the region $[6,9, 7.0]$. The right subfigure shows the simulation results of the traffic model at $t = 3$. Only one backward wave remains.}  
\end{figure}

We can facilitate simulation if we harness the theoretical aspects of linear hyperbolic equations. To simulate the first-order wave equation, only a one-side scheme is feasible and the low accuracy of the scheme requires a very dense discretization \cite{thomas2013numerical} which makes the computation task a nightmare. Thus, the simulation is just done by moving the waves along the axis. The results are shown in  Figure \ref{fig:71} and \ref{fig:72}.

At $t = 1$, the backward wave appearing at the boundary $x = 0$ is about to leave. Wave originating from $[3.5, 3.7]$ moves to the place around $2.4$. The forward wave starting at $[3.9, 4.0]$ propagates to the region $[6,9, 7.0]$. The \emph{discrete partition} $p_d$ changes to $0$ for points $\{0.3, 0.4,..., 2.3, 2.4\} \cup \{2.8, 2.9,..., 9.9, 10\}$ and $1$ for points $\{0, 0.1, 0.2\} \cup \{2.5, 2.6, 2.7\}$.
At $t = 3$, only one single backward wave remains.  $p_d = 0$ for $\{0, 0.1,..., 0.7, 0.8\} \cup [1.1, 1.2, ..., 9.9, 10] $ and $p_d = 1$ for $\{0.9, 1.0\}$. \textcolor{black}{Here we use the explicit solutions instead of numerical solutions for the ODE system due to the transport feature of the traffic model.}
Simulations show that the traffic merges around $t = 0.5$ and eventually all flows leave the system. Around the merging location, the area that used to be in the free mode switches to the congestion mode. 

%% file: conclusion.tex
\section{Conclusion}\label{sec:conclusion}

In this paper, we propose a theoretical model called PDHA for modelling PDE dynamic systems. PDHA involves novel features such as the notion of domain $X$, which is used to define the spatial location, and domain partition $P$, which is used to describe the mode occupation and domain changing with respect to time. After building the theory for PDHA, we present another model called discrete space PDHA that we construct for the purpose of analysis. In discrete space PDHA, the discretization relation and scheme are proposed to formalize the discrete model. Additionally, the partition and space mode are replaced by a discrete partition, making the analysis easier. Moreover, we formally defined concepts such as time trajectory, model execution, Zeno behavior, reachability analysis, and determinism for DSPDHA. The paper also displays explorations towards the relations of the three models. 

Although a few results and similarities are presented regarding DSPDHA and HA, other topics such as invariants, uniqueness, and continuity remain unexplored. The dynamic properties we proposed are only for DSPDHA, there is no similar conclusion for the continuous PDHA. Also, in our examples, we simply consider a one-dimension uniform discretization using easy FDMs. Other approaches like generalized finite difference and finite element methods can also play a critical role if an irregular domain is considered. 
Deciding DSPDHA from PDHA is also a challenging endeavor. Thus, an approach for measuring the distance between the approximated DSPDHA and real PDHA is deeply needed.

\textcolor{black}{ It is worth mentioning that the number of ODEs in the finite difference scheme (FDM) framework depends on the chosen partition. To obtain accurate and highly-resolved solutions, a fine partition becomes indispensable, which consequently leads to high-dimensional ODE systems. In such cases, the use of deep neural network models \cite{Chen2018} could potentially offer valuable insights and broaden the applications of our approach. Exploring the potential of deep neural networks in this context is an avenue we plan to investigate in future research.  }

\section{Acknowledgements}
W. Xiang was supported by the National Science Foundation, under NSF CAREER Award no. 2143351, and NSF CNS Award no. 2223035.